\documentstyle[aps,preprint]{revtex}
\newcommand{\spmine}{1.05}
\newcommand{\myspace}{\edef\baselinestretch{\spmine}\Large\normalsize}
\begin{document}
\myspace
%

\title{Low lying excitations of the Dynamical Jahn-Teller \\ ions
C$_{60}^-$ and C$_{60}^{2-}$}

\author{Nicola Manini,$^{1,2}$ and Erio Tosatti$^{1,2,3}$}
\address{$^1$ Istituto Nazionale di Fisica della Materia (INFM)}
\address{$^2$ International School for Advanced Studies (SISSA), Via Beirut 4,
I-34013 Trieste, Italy}
\address{$^3$ International Centre for Theoretical Physics (ICTP), P.O. Box
586, I-34014 Trieste, Italy}
%
\maketitle
\vspace{10mm}
\begin{abstract}
We compute the vibronic spectrum of gas phase fullerene negative ions
C$_{60}^{-}$ and C$_{60}^{2-}$.  We treat accurately the linear dynamical
Jahn-Teller of these ions in their ground state. In particular, coupling of
the $t_{1u}$ orbital to the eight $H_g$ vibrational modes is handled by
exact diagonalizations including up to six vibrons. The resulting spectrum
is characterized by large splittings, which should be readily observable
spectroscopically. The lowest excitation symmetry is argued to be related
to the Berry phase present in this problem.
\end{abstract}
%
%
\newpage

\section{Introduction}

Due to the partial filling of the degenerate $t_{1u}$ orbital, the
fullerene anions C$_{60}^{n-}$ are Jahn-Teller (JT)
distorted\cite{negri,vzr,sch,antr,AMT,MTA,wang}. We have recently argued
that {\em dynamical} JT phenomena must be of relevance in the ground state,
at least for the gas-phase isolated negative ions. Realistic values for the
coupling between the $t_{1u}$ electronic orbital (which is derived from a
molecular $L=5$ state, but can be treated effectively as an $L=1$ state)
and the $H_g$ vibrations (quadrupolar, $L=2$) have been estimated in
literature\cite{negri,vzr,sch,antr,AMT,MTA,wang,Gunnarsson}.

The electron-vibron coupled problem in C$_{60}$ ions is affected by a Berry
phase\cite{AMT,MTA,Assa}, which is intriguing and interesting {\em per se}.
We have shown\cite{AMT,MTA} that the presence of a Berry phase implies a
number of signatures:
\begin{enumerate}
\item{a $T_{1u}$ ($L=1$) ground state for C$_{60}^{n-}$ with odd $n$, and
with even $n$ and high spin ($S=1$), against an $A_g$ ($L=0$) ground state
for even $n$ and $S=0$;}
\label{gs:point}
\item{impossibility of s-wave electron attachment
to C$_{60}$\cite{smith,attachment} due to non-existence of $A_g$ states in
the excitation spectrum of C$_{60}^{-}$;}
\label{attachment:point}
\item{large and characteristic
splittings in the excitation spectrum of vibronic states derived from the
JT active $H_g$ modes. For $^2$C$_{60}^{-}$ (and $^3$C$_{60}^{2-}$, whose
JT energy shift and excitation spectrum is identical, due to electron-hole
symmetry) the lowest excitation is a $T_{2u} + G_u$ nearly degenerate
multiplet ($L=3$ in spherical symmetry). This is a remnant of the
ortho-hydrogen-like series $L=1,3,5,...$ predicted by the Berry phase in
the semiclassical limit\cite{AMT}. For $^1$C$_{60}^{2-}$, conversely, the
lowest excitation above the $L=0$ ground state should be $H_g$ ($L=2$ in
spherical symmetry), which in turn is the remnant of the para-hydrogen-like
series $L=0,2,4,...$, predicted semiclassically.}
\label{excitations:point}
\end{enumerate}

Of the above, points \ref{gs:point} and \ref{attachment:point} are
well-established, but very little\cite{yang,heath,gasyna} is known about
the negative ion vibronic excitations of point \ref{excitations:point}. The
scope of this paper is to make quantitative predictions on the low-lying
excitation spectrum of $^2$C$_{60}^{-}$, $^3$C$_{60}^{2-}$ and
$^1$C$_{60}^{2-}$, which could enable a future precise assessment of the
coupling constant values, and of the Berry phase effect on the dynamics of
these interesting molecular ions.  Both the C$_{60}^{-}$\cite{yang} and
C$_{60}^{2-}$\cite{limbach} ions have been shown to exist in gas phase and
vibronic spectroscopy should be feasible in the near future.  For
C$_{60}^{2-}$ one would also obtain in this way a clear indication of
whether the ground state is $S=1$ or $S=0$, since the spectra are very
different in the two cases.

\section{Calculations}

In C$_{60}^{n-}$, $n$ electron in the $t_{1u}$ state are JT-coupled to the
$H_g$ molecular vibrations. Due to extreme stiffness and harmonicity of the
C$_{60}$ cage, linear coupling provides an accurate description of this JT
problem for $n=1$ and $n=2$ ($S=1$), and a less accurate but still
qualitatively correct description for $n=2$ ($S=0$)\cite{MTA}. The
Hamiltonian describing the coupling to the eight (fivefold degenerate)
$H_g$ modes is\cite{obrien,vzr,MTA}
\begin{eqnarray}
H		&=& \sum_{k=1}^8{ H^0_k + H^{e-v}_k }\nonumber \\
H^0_k		&=&{ \hbar \omega_k \over 2} \sum_{\mu }
\left(P_{k,\mu}^2 + Q_{k,\mu}^2 \right) {\bf I} \nonumber \\
H^{e-v}_k	&=& g_k {\sqrt{3}\over 2}\hbar\omega_k \sum_\sigma \nonumber \\
&&(c^\dagger_{x\sigma}, c^\dagger_{y\sigma},c^\dagger_{z\sigma})
\pmatrix{
{1 \over \sqrt{3}} Q_0 +Q_2 &-Q_{-2}        		&-Q_1 \cr
-Q_{-2} 		&{q \over \sqrt{3}} Q_{0}-Q_{2}	&-Q_{-1}\cr
-Q_1			&-Q_{-1}	&- {2 \over \sqrt{3}} Q_{0} \cr
}					\pmatrix{       c_{x\sigma}\cr
                                                        c_{y\sigma}\cr
                                                        c_{z\sigma} }
+ ...
\label{ev_Hamiltonian}
\end{eqnarray}
where $Q_{k,\mu}$ is the normal co-ordinate of mode $k$, of spherical
component $\mu$, $P_{k,\mu}$ is its canonical conjugate momentum operator,
and $c^\dagger_{\alpha\sigma}$ are fermion operators creating an electron
in orbital $\alpha$ with spin $\sigma$.
This Hamiltonian is derived on the basis on the icosahedral symmetry of the
buckyball. The icosahedral group $I_h$ is a subgroup of the full rotation
group in three dimensions $O(3)$: when reducing the symmetry from $O(3)$ to
$I_h$, the representations $D^{(0+)}$, $D^{(1-)}$ and $D^{(2+)}$ go into
$a_g$, $t_{1u}$ and $h_g$ respectively, unsplit. The Clebsch-Gordan
coefficients for these representations of the two groups are thus the
same\cite{Cesare}. As Hamiltonian (\ref{ev_Hamiltonian}) is restricted to
$t_{1u}$ and $H_g$ states, it has actually full spherical symmetry: the
Hamiltonian for a $L=1$ ``superatomic'' electronic state interacting with
the normal vibrations of a concentric elastic sphere looks exactly the
same. Linear JT coupling hides a higher-than-icosahedral symmetry, and
yields states with degeneracies characteristic of the spherical group.
Higher order terms (indicated with dots in Eq. \ref{ev_Hamiltonian}: we
neglect them in the present treatment) would split these degeneracies into
the appropriate $I_h$ representations. For this reason, we use a hybrid
notation, and label states with icosahedral or spherical representations,
according to convenience.

The unperturbed frequencies $\omega_k$ of the eight $H_g$ modes of neutral
C$_{60}$ are experimentally well-known (see Table I). $A_g$ modes do not
split the $t_{1u}$ level, and are not included in this calculation. Shifts
of these $H_g$ frequencies from the neutral molecule are not expected to be
large, and will be neglected here. If needed, corrections can be made
using, e.g., the shifted frequencies calculated in solid state
A$_n$C$_{60}$ by Andreoni and collaborators\cite{andreoni}. The
dimensionless JT coupling parameters $g_k$ are more uncertain. Here we
shall adopt the recent values empirically extracted by
Gunnarsson\cite{Gunnarsson} by fitting photoemission spectra of C$_{60}^-$
in gas phase.

In an earlier work\cite{AMT,MTA}, we diagonalized Hamiltonian
(\ref{ev_Hamiltonian}) both approximately, using perturbation theory (where
all modes can be superposed linearly), and numerically, including however
a single vibrational mode. Subsequently Gunnarsson calculated the ground
state energy accurately, including all modes\cite{Gunnarsson}.

Here, aiming at gas phase spectroscopy, we shall diagonalize
(\ref{ev_Hamiltonian}) accurately, including all modes, and we shall obtain
in addition the lowest excitation spectrum as well, for $n$=1 and 2
electrons.

The vibron operators $Q_{k,\mu}$, $P_{k,\mu}$ in Hamiltonian
(\ref{ev_Hamiltonian}) are conveniently rewritten in terms of standard
boson operators $b_{k,\mu}$. A generic vibronic state is expanded as
\begin{equation}
\Psi=  \sum
	\epsilon_{k_1\mu_1,... k_N\mu_N, \alpha_1 \sigma_1, \alpha_n \sigma_n}
	b^\dagger_{k_1\mu_1} ... b^\dagger_{k_N\mu_N}
	c^\dagger_{\alpha_1 \sigma_1} ... c^\dagger_{\alpha_n \sigma_n}
	|0>
\label{state}
\end {equation}
where $|0>$ is the state with no vibrons and no electrons.  The sum in Eq.
\ref{state} extends to states with any number of vibrons, but a linear
e-v coupling implies decay as $\exp(-N)$ of components with $N$-vibron
states, for large enough $N$. For this reason, a truncated basis set
including all states up to $N$ vibrons gives a variational estimate of the
lowest eigenvalues, with hopefully good convergence with increasing $N$.

It is straightforward to compute the Hamiltonian matrix elements on the
truncated basis. This matrix is sparse, involving nonzero terms only
between states whose numbers of vibrons differ by exactly one. As mentioned
above, the approximate Hamiltonian (\ref{ev_Hamiltonian}) has full
rotational symmetry. We work on a basis in which the $z$-component $M$ of
total angular momentum is diagonal, selecting therefore a well defined
value of this component, usually $M=0$. (At this stage we do not try to
exploit the full rotational symmetry of (\ref{ev_Hamiltonian}), and to work
on a diagonal basis for the total angular momentum $L$.) In the set of
coupling parameters which we use (see Table I), mode $H_{g6}$ has vanishing
coupling: therefore we need only to include in the diagonalization the
remaining 7 modes.  In Table II and III, we list the size $d$ of the
truncated Hilbert space at different values of $N$, for $M=0$. The
structure of the problem is suitable for a Lanczos algorithm. Our
computational resources allow us to solve such a problem for $d<5\times
10^6$, thus including up to $N=6$ vibrons.  Tables II and III also show the
ground state JT energy gain so obtained. These values are very close, as
they should, to those found by Gunnarsson\cite{Gunnarsson}. The numbers
show a clear trend towards convergence, which is better for $n=1$. For
$n=2$ ($S=0$), the JT distortion is stronger, and convergence much more
problematic.  The convergence behavior for a few low-lying levels is
illustrated in Fig. \ref{convergence} as a function of $1/N^2$: panel (a)
refers to $n=1$ (and equivalently to the $n=2$, $S=1$), while panel (b) is
for $n=2$, $S=0$. We have indicated a polynomial interpolation (continuous
curves) simply as a guide to the eye. We have restricted this calculation
to the lowest 5-6 eigenvalues, covering excitation energies only up to
400-500 cm$^{-1}$. This is the region where the results are more reliable,
within the limits imposed by the residual uncertainties in the coupling
constants $g_k$. At higher excitation energies, multi-mode combinations
make the spectrum increasingly dense, and unreliable. Moreover, the Lanczos
algorithm, as implemented, is unsuitable for computing accurately a large
number of excited levels.

\section{Results and discussion} \label{results}

The predicted excitation spectrum of gas phase $^2$C$_{60}^-$ and
$^3$C$_{60}^{2-}$ within the linear coupling spherical approximation) is
shown in Fig. \ref{spectrum} (a), and is reported in Table II.  The four
lowest excitations are essentially derived from the $H_{g 1}$ mode at
270cm$^{-1}$, which is split in a qualitatively similar manner as predicted
by perturbation theory\cite{MTA}. The numerical value of the splitting
which we find here is rather large, and should be readily observable in
future Raman or EELS spectroscopy, if feasible, in gas phase. The lowest
excitation is confirmed to be the $T_{2u} + G_u$ ($L=3$) doublet near 200
cm$^{-1}$. As discussed above this doublet will in reality be split by
higher order, nonspherical terms, absent in the Hamiltonian
(\ref{ev_Hamiltonian}).  However, we expect that the nonspherical splitting
should be rather small relative to that to the next $H_u$ state ($\approx$
70 cm$^{-1}$). Experimental confirmation of the overall ``$L=3$'' nature of
the lowest excitation doublet would provide an interesting check for the
Berry phase in C$_{60}^-$.  Below 300 cm$^{-1}$ the remaining excitations
($H_u$, $T_{1u}$) retain a prevalent $\nu_1$ character.

Above the $\nu_1$-derived multiplet, the next $L=3$ excitation, near 357
cm$^{-1}$, has largely $\nu_2$ character, but there is also admixture with
$2\nu_1$ components. We note that there are no low-lying $L=0$ states.  It
is interesting that among all $N$-vibron states, only those which involve
{\em different} $H_g$ modes can give rise to an overall $L=0$ state. The
lowest $L=0$ state is expected to originate from the coupling with states
belonging to $\nu_1+\nu_2$, while none is originated from a pure overtone,
such as $2\nu_1$. The reason for this is the following. Since the bare
electronic state is $L=1$, we need another $L=1$ (of vibrational nature) to
get an overall $L=0$. In the combinations of two different $L=2$ vibrons,
$L=4,3,2,1,0$ states are generated, and these contain an $L=1$ ($T_{1g}$)
state. For two (or more) {\em identical} vibrons, however, only $L=4,2,0$
are generated, the odd states vanishing due to permutation symmetry. Hence,
there is no $L=1$ in the overtones, and no overall $L=0$.  We expect,
accordingly, the lowest $L=0$ state at $\approx$700 cm$^{-1}$, close to the
C$_{60}$ $\nu_1 + \nu_2$ origin.  In the icosahedral group, however this
kind of state will have $A_u$ symmetry. Conversely, $A_g$ states can only
be generated by involving ``$u$'' vibrons. But these are linearly uncoupled
to the electronic state, and thus very hard to excite. Thus it is confirmed
that an approaching s-wave electron cannot efficiently attach to the
C$_{60}$ molecule and end up (at least for the $t_{1u}$ orbital) in a global
$A_g$ ($L=0$) state\cite{attachment}.

The predicted low-lying excitation spectrum of a hypothetical gas phase
$^1$C$_{60}^{2-}$ is given in Fig. \ref{spectrum} (b), and in Table III.
Convergence is in this case much more problematic than for $n=1$, or $n=2$,
$S=1$.  There is a new feature, consisting in a crossing of levels (of
different symmetry), upon increasing $N$ (Fig. \ref{convergence} (b)). For
these reasons, error-bars on the extrapolated values are larger, probably
exceeding 10 cm$^{-1}$.

In contrast to the $n=1$ case discussed above, for $n=2$, $S=0$ not all the
lowest excitations can be classified as derived mainly from the lowest
vibrational mode $H_{g 1}$. The 6 two-electron $S=0$ states with no vibrons
are split by e-v coupling into $^1|L=0> \oplus ^1|L=2>$ vibronic
states. The singlet ground state is $^1|L=0>$, and the $^1|L=2>$ state is
left as a low-lying excitation. In the weak coupling limit its energy goes
to zero as $\sum g_k^2\hbar\omega_k$ (see Fig.5 in Ref.\cite{AMT}). For
finite coupling, contributions to this excitation wave function come
simultaneously from all $H_g$ modes, and are not dominated by lowest mode
$H_{g1}$, unlike, e.g., the lowest $L=3$ excitation of $^3$C$_{60}^{2-}$.
If $H_{g 1}$ were the only mode coupled to the $t_{1u}$ orbital, then the
predicted excitation energy for this lowest $L=2$ state, would be 69.9
cm$^{-1}$.  The energy we find when all the modes are included (Table III)
is significantly smaller, approximately 40cm$^{-1}$, confirming the
participation of all modes.

The next higher excitations are an $L=4$ state (unsplit doublet $G_g+H_g$),
followed by an $L=6$, $L=2$ and $L=3$, all essentially of $H_{g 1}$ origin.
This sequence constitutes a clear remnant of the para-hydrogen-like series
$L=0,2,4,6,...$ predicted semiclassically when the Berry phase cancels, as
in this case.  The first level originating from $H_{g2}$ is, as expected, a
state of symmetry $L=4$ around 330 cm$^{-1}$ above the ground state,
identically beginning another series $L=4,6,8,...$. It should be noted here
that icosahedral splittings (not included) are probably larger in this case
relative to $n=1$.

The pure JT energetics (see Table II and III) indicates the spin singlet
configuration as the preferred ground state of C$_{60}^{2-}$. However our
calculation does not include Coulomb e-e repulsion, which is not negligible
in this state. There is in fact some experimental evidence\cite{dubois}
that C$_{60}^{2-}$ ions in a matrix may prefer an $S=1$ configuration, as
expected from Hund's rule. The reduction of screening in vacuum could imply
an even stronger Coulomb repulsion, further favoring an $S=1$ ground
state. It should be noted however that large $U$ Hubbard models, as well as
PPP calculations appear to favor a singlet ground state for $n=2$, even
without JT effect\cite{Baskaran,negri2}. On the other hand, the dynamical
JT effect of the ion when in a matrix is likely to be quenched (by crystal
fields or by polarization of the surrounding medium) to a static JT
distorsion, which gains considerably less energy than the dynamic JT effect
considered here\cite{Gunnarsson}. This leaves open a residual chance of a
$^1$C$_{60}^{2-}$ ground state in gas phase.

We stress here finally, that our spectra have been computed neglecting
spin-orbit effects. Since the intra-carbon spin-orbit coupling is
$\zeta(2p_c)\approx 28$ cm$^{-1}$\cite{mcglynn}, it can be expected to
yield additional splittings, both in the ground state and the low lying
excitations, of comparable magnitude. Splittings of 30 and 75 cm$^{-1}$
reported in the near IR spectrum of C$_{60}^-$ in solid Ar have in fact
been attributed to spin-orbit\cite{gasyna}. Interpretation of the actual
spectra of C$_{60}^-$ and C$_{60}^{2-}$ will have to include consideration
of spin-orbit effects, to be calculated in the future.

\section{Conclusions}

We have calculated the lowest excitation spectrum, derived by JT coupling
of the $H_g$ modes for C$_{60}^-$ and C$_{60}^{2-}$ in gas phase. The
calculation involves neglecting nonspherical and nonlinear effects, as well
as spin-orbit coupling, but is otherwise fully quantitative. Common
characters of these spectra are sizable splittings, in the order of several
tens of cm$^{-1}$, and a substantial lowering of the transition energy to
the lowest mode, in comparison with the lowest $H_g$ mode of neutral
C$_{60}$. The symmetry of this lowest excitation, in particular, should be
revealing in connection with the Berry phase which affects this Jahn-Teller
problem\cite{Assa,AMT}.
Gas phase vibron spectroscopy of these negative ions will be very important
in the future, to check these predictions.

\section*{Acknowledgements}
It is a pleasure to thank G.  Santoro, M. Fabrizio, P. Br\"{u}hwiler,
O. Gunnarsson, G. Orlandi and L. Yu for discussions. E. Tosatti
acknowledges the sponsorship of NATO, through CRG 920828.


\vskip 2.5cm
\begin{center}
\begin{tabular}{ccc} \hline \hline
$H_{gk}$	&$\hbar\omega_k$[cm$^{-1}$]&$g_k$\\
\hline
1	&	270.0	&	.868	\\
2	&	430.5	&	.924	\\
3	&	708.5	&	.405	\\
4	&	772.5	&	.448	\\
5	&	1099.0	&	.325	\\
6	&	1248.0	&	.000	\\
7	&	1426.0	&	.368	\\
8	&	1575.0	&	.368	\\
\end{tabular}
\end{center}
\vskip 0.1in
TABLE I. The vibrational frequencies and couplings for the JT-active $H_g$
modes of fullerene.
\newpage

\begin{center}
\begin{tabular}{ccccccccc} \hline \hline
$N$&$d$	&$L=1$  	&$L=3$	&$L=2$	&$L=1$	&$L=3$	&$L=5$	&$L=3$	\\
\hline
3&3662	&-1094.6	&-810.2	&-742.2	&-721.3	&-657.1	&-350.8	&-314.8	\\
4&31459	&-1117.8	&-885.6	&-814.3	&-793.6	&-729.5	&-588.3	&-548.2	\\
5&225716&-1124.2	&-910.1	&-837.8	&-818.5	&-753.1	&-665.7	&-625.0	\\
6&1405972&-1125.7	&-917.0	&-844.3	&-826.2	&-759.7	&-691.1	&-650.5	\\
7&7775140&-		&-	&-	&-	&-	&-	&-	\\
$\infty$&&-1126.3	&-926.7	&-853.5	&-838.9	&-768.8	&-728.6	&-689.3	\\
\end{tabular}
\end{center}
\vskip 0.1in
TABLE II. Low lying levels obtained by exact diagonalization of Hamiltonian
(\ref{ev_Hamiltonian}) C$_{60}^-$ or for C$_{60}^{2-}$ ($S=1$) for
increasing vibron number $N$. Symmetries of these vibronic states are
indicated at the top. The extrapolated $N\to\infty$ energies (see figure
\ref{convergence}) are at bottom. The energy zero is the undistorted
molecule, and energies are in cm$^{-1}$.
\vskip 2.5cm

\begin{center}
\begin{tabular}{ccccccccc} \hline \hline
$N$&$d$		&$L=0$		&$L=2$		&$L=4$		&$L=6$		&$L=2$		&$L=3$		&$L=4$\\
\hline
3&10650	&	-2898.3	&	-2643.7	&	-2075.7	&	-947.3	&	-2278.4	&	-1880.9	&	-1840.8	\\
4&91689	&	-3085.2	&	-2907.7	&	-2547.7	&	-1934.1	&	-2591.9	&	-2343.4	&	-2310.8
\\
5&660936&	-3201.0	&	-3068.2	&	-2808.2	&	-2409.9	&	-2778.9	&	-2606.9	&	-2574.8
\\
6&4130556&	-3274.0	&	-3167.8	&	-2964.6	&	-2670.1	&	-2896.3	&	-2768.6	&	-2735.2
\\
7&22907016&		&-		&-		&-		&-		&-		&-		\\
$\infty$&&	-3459.5	&	-3419.9	&	-3348.2	&	-3275.9	&	-3190.9	&	-3168.9	&	-3130.6
\\
\end{tabular}
\end{center}
\vskip 0.1in
TABLE III. Low lying levels obtained by exact diagonalization of
Hamiltonian (\ref{ev_Hamiltonian}) for C$_{60}^{2-}$ ($S=0$) for increasing
vibron number $N$. Symmetries are indicated at the top. The extrapolated
energy (see figure \ref{convergence}) is at bottom. As discussed in the
text, the uncertainty on these extrapolated energies is rather large.  The
energy zero is the undistorted molecule, and energies are in cm$^{-1}$.


\begin{figure}
\caption{The C$_{60}^{n-}$ ground state and low excitation energies, (a)
for $n=1$ and $n=2$ $S=1$, and (b) for $n=2$ $S=0$, as a function $1/N^2$
($N$ is the number of vibrons included in the calculation).  Continuous
lines are fits for $N>3$. }
\label{convergence}
\end{figure}

\begin{figure}
\caption{Extrapolation for $N\to\infty$ of the C$_{60}^{n-}$
excitation spectra, (a) for $n=1$ and $n=2$ $S=1$, and (b) for $n=2$
$S=0$. The symmetry species of the levels are indicated. All
non-Jahn-Teller excitations are omitted.}
\label{spectrum}
\end{figure}

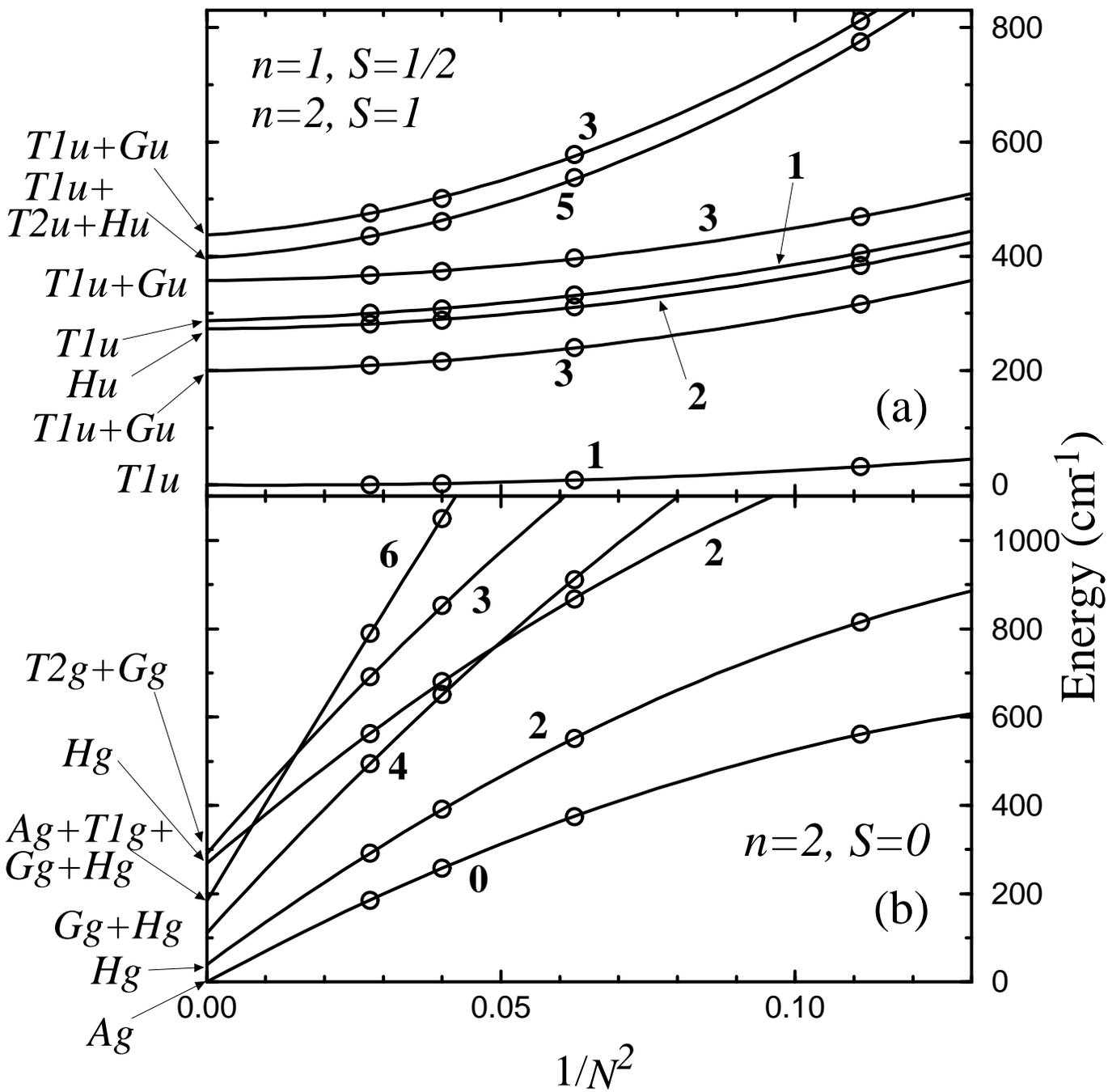
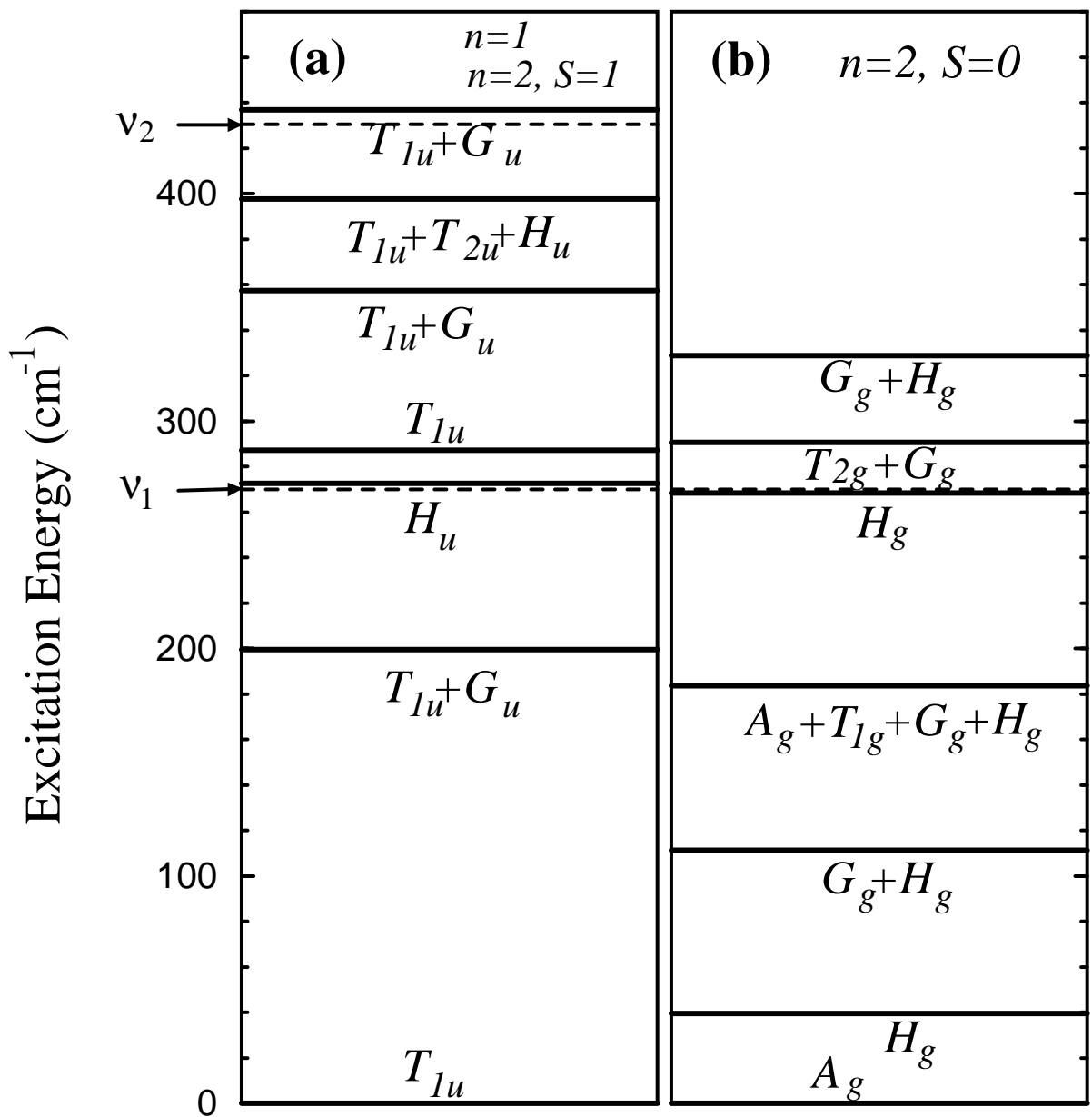
\end{document}